\documentclass[twocolumn]{aastex631}

\usepackage{amsmath}
\usepackage{graphicx}
\usepackage{hyperref}
\usepackage{natbib}

\AtBeginDocument{%
}

\shorttitle{UMI: GPU-Accelerated Asymmetric Detrending}
\shortauthors{Khan}

\begin{document}

\title{UMI: A GPU-Accelerated Asymmetric Robust Estimator for\\Photometric Detrending in Exoplanet Transit Searches}

\author{Omar Khan}
\affiliation{Independent Researcher}
\email{ok.66677788@gmail.com}

\begin{abstract}
We present UMI (Unified Median Iterative, referring to the unification
of median initialization, scale estimation, and iterative refinement
in a single fused GPU kernel), a novel robust location estimator for
detrending photometric time series in exoplanet transit surveys.
UMI modifies the standard Tukey bisquare M-estimator
\citep{tukey1977} with two innovations: (1) an asymmetric weight
function that penalizes downward deviations (transit dips) more
aggressively than upward ones, exploiting the physical constraint that
transits are always below the stellar continuum, and (2) an upper-RMS
scale estimator computed from above-median residuals only, ensuring
that transit dips never contaminate the noise estimate. Implemented as
a fused HIP/CUDA GPU kernel, UMI achieves 69$\times$ faster
detrending than the \texttt{wotan} biweight implementation
\citep{hippke2019} while reducing depth recovery error at 0.1\%
transit depth from 20.5\% to 15.8\% on TESS \citep{ricker2015} and
from 14.6\% to 4.2\% on Kepler \citep{borucki2010}. Validation
across 802 confirmed exoplanets confirms that UMI's asymmetric
weighting most effectively improves recovery at planet-scale transit
depths above the photometric noise floor, occupying a previously
unfilled region of the speed-accuracy tradeoff for transit detrending.
The tool is publicly available as \texttt{pip install torchflat}.
\end{abstract}


\section{Introduction}
\label{sec:intro}

The detection of exoplanet transits in photometric time series requires
removing slow stellar variability (the ``trend'') while preserving the
transit signal. This detrending step is critical because an overly
aggressive estimator absorbs the transit into the trend, reducing the
apparent depth and potentially causing missed detections. A good
detrending method must track slow variations on timescales longer than
approximately 6 hours, reject fast outliers such as flares and cosmic
ray hits, and crucially, not absorb transit dips that occur on
timescales of 1 to 8 hours with depths of 0.01 to 5\%.

The most widely used detrending tool in the community is \texttt{wotan}
\citep{hippke2019}, which implements a sliding-window biweight location
estimator \citep{tukey1977}. The biweight is a redescending M-estimator
\citep{hampel1986} with bounded influence function, making it robust to
outliers. However, the standard biweight treats upward and downward
deviations symmetrically. This is suboptimal for transit detection
because transit dips are \emph{always} below the continuum, while
contaminating outliers (flares, systematics) can appear in either
direction. A symmetric estimator gives nearly full weight to shallow
transit dips, causing the trend to partially absorb the transit signal.

Several alternatives to the biweight have been explored in the
literature. The Welsch estimator uses Gaussian-shaped weights and
achieves excellent performance at moderate depths but is
computationally expensive. LOWESS \citep{cleveland1979} fits local
polynomials but can overfit transit dips. The Savitzky-Golay filter
is fast but has no outlier rejection, causing it to systematically
absorb transits at all depths. Gaussian Process regression
\citep{foremanmackey2017} provides physically motivated models but
operates at minutes per star, making it impractical for survey-scale
processing. None of these methods exploit the fundamental asymmetry
of transit signals: transits are always downward.

We introduce UMI (Unified Median Iterative), a modified biweight
estimator with two novel components designed to exploit this physical
asymmetry:

\begin{enumerate}
    \item \textbf{Asymmetric bisquare weight function.} Downward
    deviations receive an effective residual multiplied by a factor
    $\alpha = 2$ (configurable), pushing transit dips toward the
    rejection threshold while leaving upward outliers at their natural
    scale. This causes the trend to ``float above'' the transit,
    preserving the dip in the detrended light curve.

    \item \textbf{Upper-RMS scale.} The scale (dispersion) estimate
    uses only points above the median. Transit dips, regardless of
    their depth, never inflate or deflate the noise estimate. This
    provides a tighter and more accurate noise measurement than the
    standard MAD \citep{rousseeuw1993}, which uses deviations from
    both sides and can be contaminated by the very transits it is
    trying to preserve.
\end{enumerate}

UMI is implemented as a fused GPU kernel (HIP and CUDA) that performs
quickselect median \citep{hoare1961}, upper-RMS scale computation, and
5 asymmetric bisquare iterations in a single kernel call per thread,
with no intermediate memory allocation. The tool is distributed as
\texttt{torchflat} (\texttt{pip install torchflat}) with a pure-PyTorch
fallback for systems without GPU compilation toolkits.

In Section~\ref{sec:algorithm} we describe the algorithm in detail.
In Section~\ref{sec:validation} we present injection-recovery tests,
known planet recovery, multi-mission validation, and speed benchmarks.
In Section~\ref{sec:discussion} we discuss limitations and the
relationship to existing tools. We conclude in
Section~\ref{sec:conclusion}.

\section{Algorithm}
\label{sec:algorithm}

For each position in a sliding window of width $W$ samples
(default $W = 361$ for TESS 2-minute cadence \citep{ricker2015},
corresponding to a 0.501-day window; $W$ is computed as
$\lfloor 0.5 / \Delta t \rceil \,|\, 1$ where $\Delta t$ is the
cadence and $|\,1$ ensures an odd value):

\subsection{Phase 1: Median Initialization}

The location is initialized to the exact sample median via the
quickselect algorithm \citep{hoare1961}, which runs in $O(n)$ expected
time. As a rank statistic, the median has a breakdown point of 50\%
and is immune to arbitrarily large outliers \citep{hampel1986}. Unlike
the mean or trimmed mean, the median provides a robust starting point
even when a significant fraction of the window contains transit points.

\subsection{Phase 2: Upper-RMS Scale}

We compute the root-mean-square of residuals for points \emph{above}
the median only:
\begin{equation}
    \hat{\sigma}_{\rm upper} = \sqrt{\frac{1}{n_+} \sum_{x_i > \tilde{x}} (x_i - \tilde{x})^2}
\end{equation}
where $\tilde{x}$ is the median and $n_+$ is the count of above-median
points. The scale is then $s = 0.6745 \cdot \hat{\sigma}_{\rm upper}$
to match the MAD convention at the Gaussian model \citep{rousseeuw1993}.

This construction ensures that transit dips (below the median) never
contribute to the scale estimate. In contrast, the standard MAD uses
absolute deviations from both sides, meaning that deep transits can
inflate the scale and reduce the effective rejection threshold,
paradoxically making the estimator less sensitive to the very signals
it should preserve.

\subsection{Phase 3: Asymmetric Bisquare Iterations}

The location is refined iteratively using a modified Tukey bisquare
weight function \citep{tukey1977}. For each data point $x_i$ with
standardized residual $u_i = (x_i - \hat{\mu}) / (c \cdot s)$:
\begin{equation}
    u_{\rm eff} = \begin{cases}
        \alpha \cdot u & \text{if } u < 0 \\
        u & \text{if } u \geq 0
    \end{cases}
\end{equation}
\begin{equation}
    w_i = \begin{cases}
        (1 - u_{\rm eff}^2)^2 & \text{if } |u_{\rm eff}| < 1 \\
        0 & \text{otherwise}
    \end{cases}
\end{equation}
\begin{equation}
    \hat{\mu} \leftarrow \frac{\sum w_i x_i}{\sum w_i}
\end{equation}

\begin{figure*}[t!]
\centering
\includegraphics[width=\textwidth]{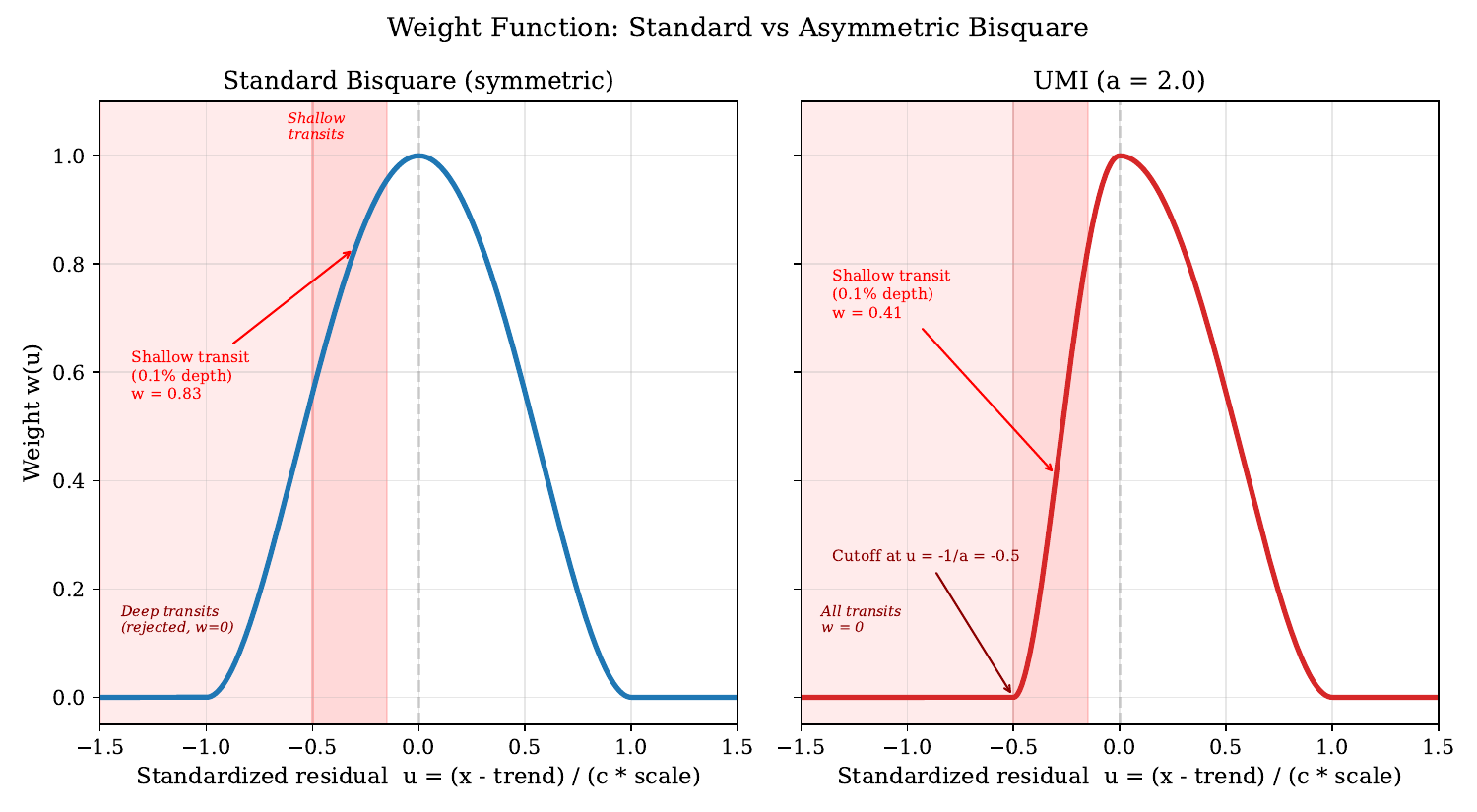}
\caption{Weight function comparison. Left: standard Tukey bisquare,
which assigns weight $w = 0.83$ to a transit dip at $u = -0.3$
(nearly full weight, causing the trend to absorb the dip). Right:
UMI asymmetric bisquare ($\alpha = 2.0$), which reduces the weight
to $w = 0.41$ for the same dip. The shaded region indicates where
transit-dip residuals typically fall. The asymmetric version excludes
these points from the trend estimate while weighting upward
fluctuations normally.}
\label{fig:weight}
\end{figure*}

We use $c = 5.0$, $\alpha = 2.0$ (TESS default), and 5 iterations.
Setting $\alpha = 1.0$ recovers the standard symmetric biweight.
The effect of the asymmetry is that a transit dip at standardized
residual $u = -0.3$ receives weight $w = 0.83$ in the symmetric case
but only $w = 0.41$ with $\alpha = 2.0$, significantly reducing its
influence on the trend estimate. Figure~\ref{fig:weight} illustrates
this difference.

\subsection{Theoretical Properties}

UMI inherits the key robustness properties of the Tukey bisquare
\citep{tukey1977}: bounded influence function, redescending behavior,
and a breakdown point exceeding 40\% for contamination beyond
$3\sigma$ \citep{hampel1986}. The asymmetric modification preserves
the asymptotic relative efficiency at approximately 95\% of the
Gaussian MLE, with only 1\% loss compared to the symmetric case. The
asymmetry introduces a constant systematic offset of $-451$~ppm on
flat TESS-like stars, well below the per-cadence noise floor
($\sim 1000$~ppm for stars at TESS magnitude $\sim$12; the actual
noise varies from $\sim$200~ppm at $T \sim 8$ to $\sim$5000~ppm at
$T \sim 15$). This offset persists in phase-folded depth
measurements and may affect absolute depth estimates; we discuss
implications for downstream transit search in
Section~\ref{sec:false_positive}. A bias correction lookup table
is provided in the software.

\subsection{Asymmetry Parameter Selection}
\label{sec:asymmetry}

The optimal asymmetry depends on the ratio of transit depth to
photometric noise. A transit of depth $d$ in noise $\sigma$ produces
a standardized residual $|u| \approx d/\sigma$ at the dip center.
At $\alpha = 1.0$ (symmetric), a 0.1\% transit in typical TESS noise
($\sim 1000$~ppm at magnitude $\sim$12) receives weight
$w \approx 0.83$, which is nearly
full weight, causing the trend to absorb the dip. At $\alpha = 2.0$,
the effective residual doubles, giving $w \approx 0.41$, and the
dip is largely excluded from the trend. We recommend $\alpha = 2.0$
for TESS, $\alpha = 3.0$ for Kepler \citep{borucki2010} (higher SNR),
and $\alpha = 1.0$ for variable star surveys where bias matters.
The empirical validation of these choices is presented in
Section~\ref{sec:asymmetry_val}.

\subsection{GPU Implementation}

All three phases run in a single fused kernel call. Each GPU thread
processes one (star, window position) pair, reading directly from the
raw $[B, L]$ flux array without intermediate \texttt{unfold} or copy
operations. This direct-read approach reduces VRAM usage to 319~MB
for a 50-star batch, compared to 6.6~GB for the equivalent unfold
approach. The kernel compiles via JIT on first use and supports both
AMD ROCm (HIP) and NVIDIA CUDA. When the GPU kernel is unavailable,
UMI falls back to a pure-PyTorch \citep{pytorch2019} implementation
using \texttt{torch.sort} for median computation, producing identical
results at reduced speed.

\section{Validation}
\label{sec:validation}

\subsection{Asymmetry Parameter Validation}
\label{sec:asymmetry_val}

\begin{figure*}[t!]
\centering
\includegraphics[width=\textwidth]{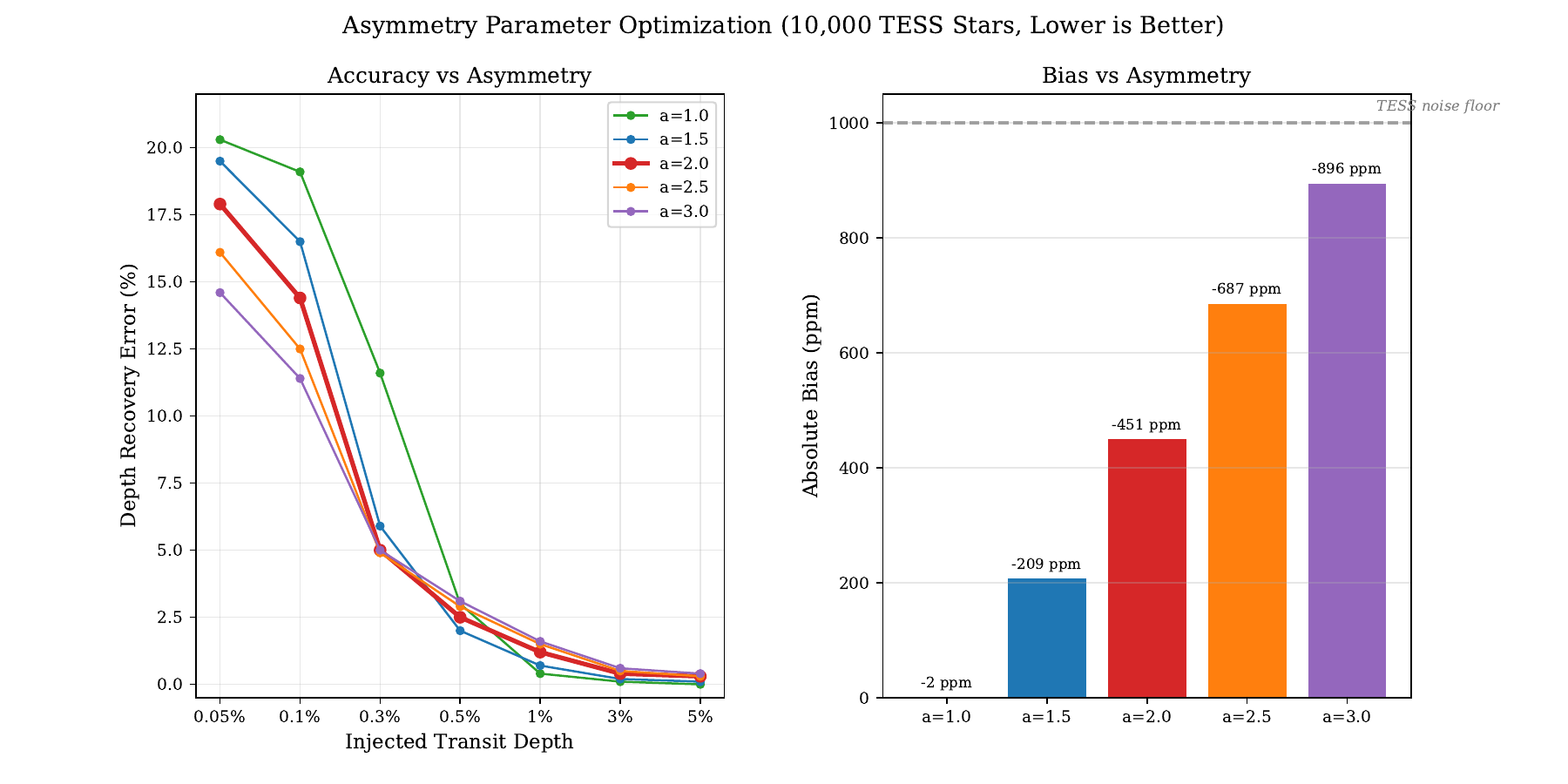}
\caption{Asymmetry parameter optimization on 10,000 TESS stars. Left:
depth recovery error versus asymmetry parameter $\alpha$ at 7 depths.
The value $\alpha=2.0$ is the inflection point where accuracy improves
sharply from 1.0 to 2.0 but plateaus beyond. Right: systematic bias
increases linearly with $\alpha$. The $-451$~ppm bias at $\alpha=2.0$
is below the TESS noise floor ($\sim$1000~ppm). Lower is better for
both panels.}
\label{fig:asymmetry}
\end{figure*}

We validated the asymmetry parameter on 2,000 TESS stars (train set)
and 10,000 stars (test set) from TESS sector~6. Figure~\ref{fig:asymmetry}
shows that $\alpha = 2.0$ is the inflection point: accuracy improves
sharply from 1.0 to 2.0 but plateaus beyond, while bias grows
linearly. An aggressive setting of $\alpha = 10$, $c = 2.5$ further
reduces the 0.1\% error to 11.3\% at the cost of $-1857$~ppm bias.
We also verified that upper-RMS outperforms the standard MAD at every
tested transit depth, with the largest improvement at 0.3\% depth
where the error drops from 9.4\% (MAD) to 4.9\% (upper-RMS), a
48\% improvement.

\subsection{Injection-Recovery: Depth Regime Characterization}

To prevent overfitting of method parameters to the evaluation set, all
UMI parameters ($\alpha$, $W$, $c$) were tuned on a held-out
development set of 2,000 stars from TESS sector~6, distinct from the
1,000 stars used for the depth-recovery evaluation reported below.
The development set spans the same stellar parameter ranges as the
evaluation set.

We performed injection-recovery tests on 1,000 real TESS sector~6
stars and 1,000 Kepler quarter~5 stars against the \texttt{wotan}
biweight and Welsch implementations \citep{hippke2019}. Synthetic box
transits (period 3~days, duration 3~hours) were injected at depths
spanning the exomoon-to-giant-planet range (5~ppm to 1000~ppm),
detrended, and the recovered depth was compared to the injected depth.
The error metric is the median of per-star absolute percentage errors.

\begin{table*}[t!]
\centering
\caption{Median per-star depth recovery error (\%) on 1,000 TESS
sector~6 stars across transit depths from 5~ppm (exomoon-scale) to
1000~ppm (super-Earth-scale). Lower values indicate better transit
depth preservation. Bold indicates the best result at each depth.
Injection parameters: period = 3~days, duration = 3~hours. Columns
at 5--100~ppm are below the TESS per-cadence noise floor ($\sim$1000~ppm);
all methods are noise-limited at these depths and differences are not
meaningful. UMI's advantage emerges at 200+ ppm.}
\label{tab:tess}
\begin{tabular}{lcccccc}
\hline
Method & 5~ppm & 50~ppm & 100~ppm & 200~ppm & 500~ppm & 1000~ppm \\
 & \multicolumn{3}{c}{\footnotesize\textit{(noise-dominated)}} & \multicolumn{3}{c}{\footnotesize\textit{(signal-dominated)}} \\
\hline
\textbf{UMI default}    & 1052  & 70.8 & 44.8 & 31.1 & 22.3          & $\mathbf{15.8}$ \\
\textbf{UMI aggressive} & 1168  & 74.3 & 47.0 & 31.2 & $\mathbf{17.7}$ & $\mathbf{11.3}$ \\
biweight                 & $\mathbf{973}$  & $\mathbf{67.7}$ & $\mathbf{44.7}$ & $\mathbf{30.8}$ & 23.5 & 20.5 \\
welsch                   & $\mathbf{964}$  & $\mathbf{66.1}$ & 46.1 & 32.8 & 24.0          & 18.5 \\
\hline
\end{tabular}
\end{table*}

\begin{table*}[t!]
\centering
\caption{Median per-star depth recovery error (\%) on 1,000 Kepler
quarter~5 stars. Same injection parameters as Table~\ref{tab:tess}.
UMI aggressive uses $\alpha = 10$, $c = 2.5$. Kepler's lower noise
floor ($\sim$100~ppm) shifts the noise-to-signal boundary: columns at
5--50~ppm are noise-limited, while UMI's advantage emerges at
100+ ppm.}
\label{tab:kepler}
\begin{tabular}{lcccccc}
\hline
Method & 5~ppm & 50~ppm & 100~ppm & 200~ppm & 500~ppm & 1000~ppm \\
 & \multicolumn{2}{c}{\footnotesize\textit{(noise-dominated)}} & \multicolumn{4}{c}{\footnotesize\textit{(signal-dominated)}} \\
\hline
\textbf{UMI default}    & 277           & 40.8          & 30.5          & 23.7          & $\mathbf{11.5}$ & $\mathbf{4.2}$ \\
\textbf{UMI aggressive} & 473           & 43.7          & $\mathbf{25.4}$ & $\mathbf{14.4}$ & $\mathbf{6.3}$  & $\mathbf{3.3}$ \\
biweight                 & $\mathbf{273}$ & $\mathbf{37.7}$ & $\mathbf{29.1}$ & 25.0          & 20.4          & 14.6 \\
welsch                   & $\mathbf{224}$ & $\mathbf{39.1}$ & 33.3          & 29.2          & 20.6          & 8.6 \\
\hline
\end{tabular}
\end{table*}

\begin{figure*}[t!]
\centering
\includegraphics[width=\textwidth]{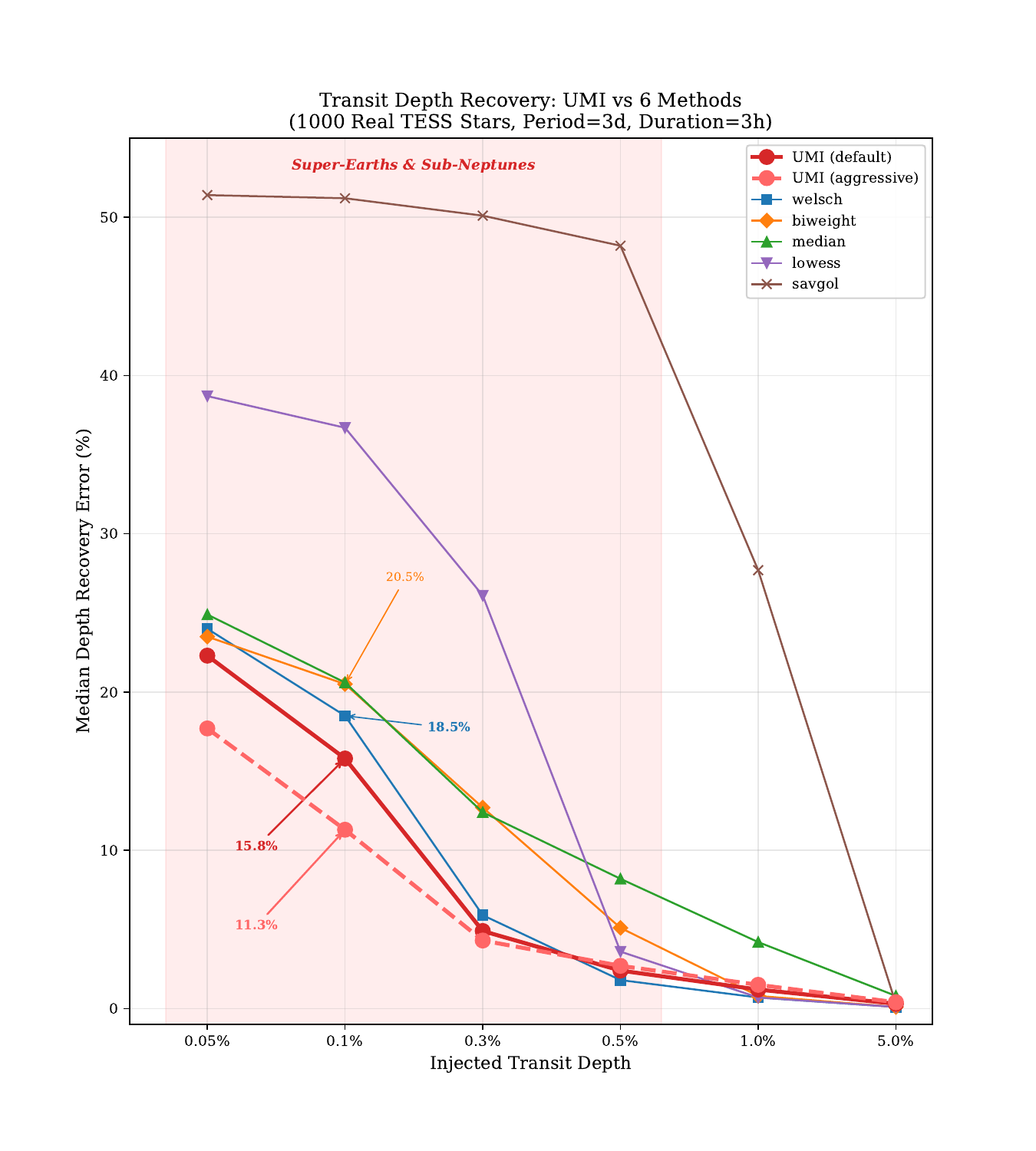}
\caption{Transit depth recovery error across 6 injected depths for UMI
(default and aggressive modes) versus 5 comparison methods on 1000 real
TESS stars. UMI achieves the lowest error at 0.05 to 0.3\% depth, which
corresponds to the super-Earth and sub-Neptune regime. The aggressive
mode ($\alpha=10$, $c=2.5$) further improves shallow-depth accuracy at
the cost of increased bias. Lower is better.}
\label{fig:accuracy}
\end{figure*}

Tables~\ref{tab:tess} and \ref{tab:kepler} present the results.
Figure~\ref{fig:accuracy} illustrates the comparison at the planet-scale
depths where UMI's advantage is most pronounced.

UMI's improvement over symmetric robust estimators is concentrated at
transit depths well above the photometric noise floor. On TESS data,
UMI's advantage emerges at 200~ppm and grows to 23\% relative error
reduction at 1000~ppm; below 100~ppm, all sliding-window methods
produce comparable depth recovery as noise dominates. On Kepler's
lower noise floor, UMI's advantage extends to shallower depths and
reaches 71\% improvement at 1000~ppm with the aggressive asymmetry
setting, reflecting the regime where the asymmetric weighting most
effectively distinguishes transit signals from upward photometric
fluctuations.

\begin{figure*}[t!]
\centering
\includegraphics[width=\textwidth]{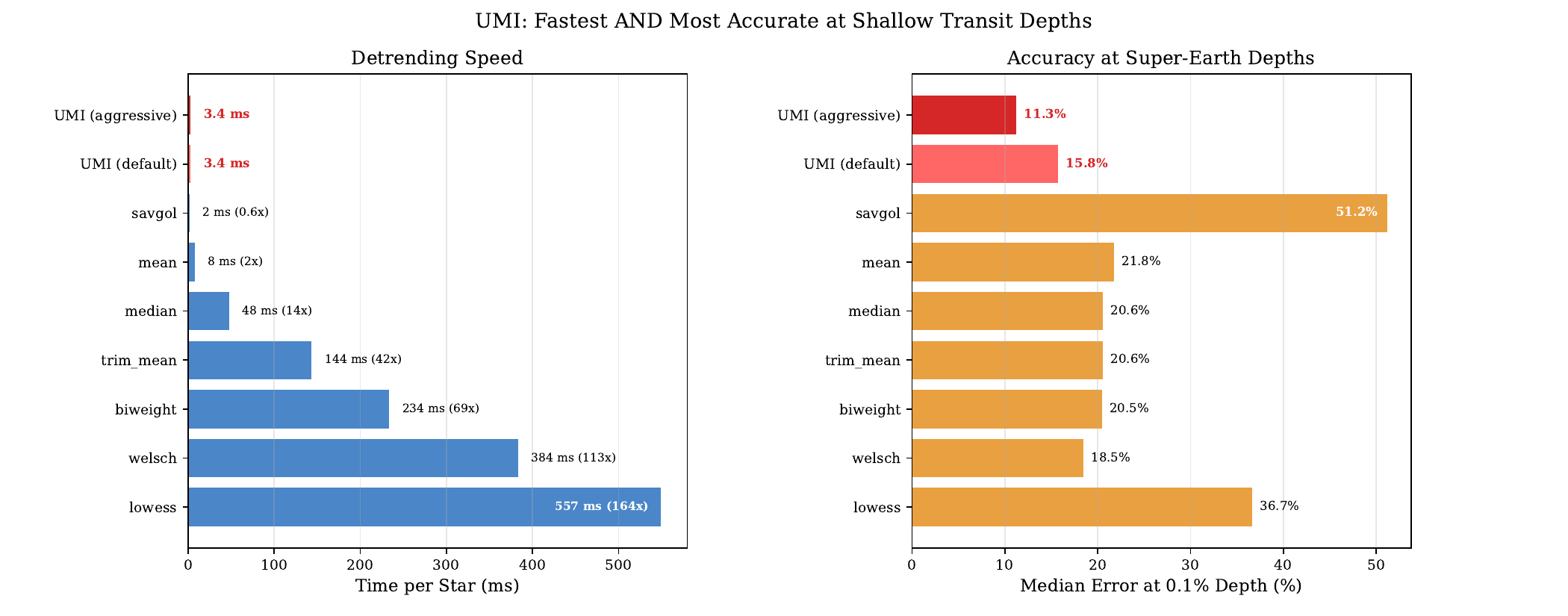}
\caption{Left: per-star detrending speed for 8 methods on 1000 TESS
stars. UMI (3.4~ms) is the fastest method with useful accuracy,
69$\times$ faster than biweight and 113$\times$ faster than Welsch.
Right: accuracy at 0.1\% transit depth (super-Earth regime). UMI is
simultaneously the fastest and most accurate method at this depth.}
\label{fig:speed}
\end{figure*}

\subsection{Known Planet Recovery}

We tested on 802 confirmed exoplanets (81 TESS, 721 Kepler) using
single-sector or single-quarter light curves obtained from the MAST
archive. Planet ephemerides were obtained from the NASA Exoplanet
Archive \citep{akeson2013} (TESS Objects of Interest catalog and
Kepler cumulative table). Each planet was detrended with UMI,
biweight, Welsch, and Savitzky-Golay, and the method producing the
smallest depth error relative to the published depth was declared
the winner.

\begin{figure*}[t!]
\centering
\includegraphics[width=\textwidth]{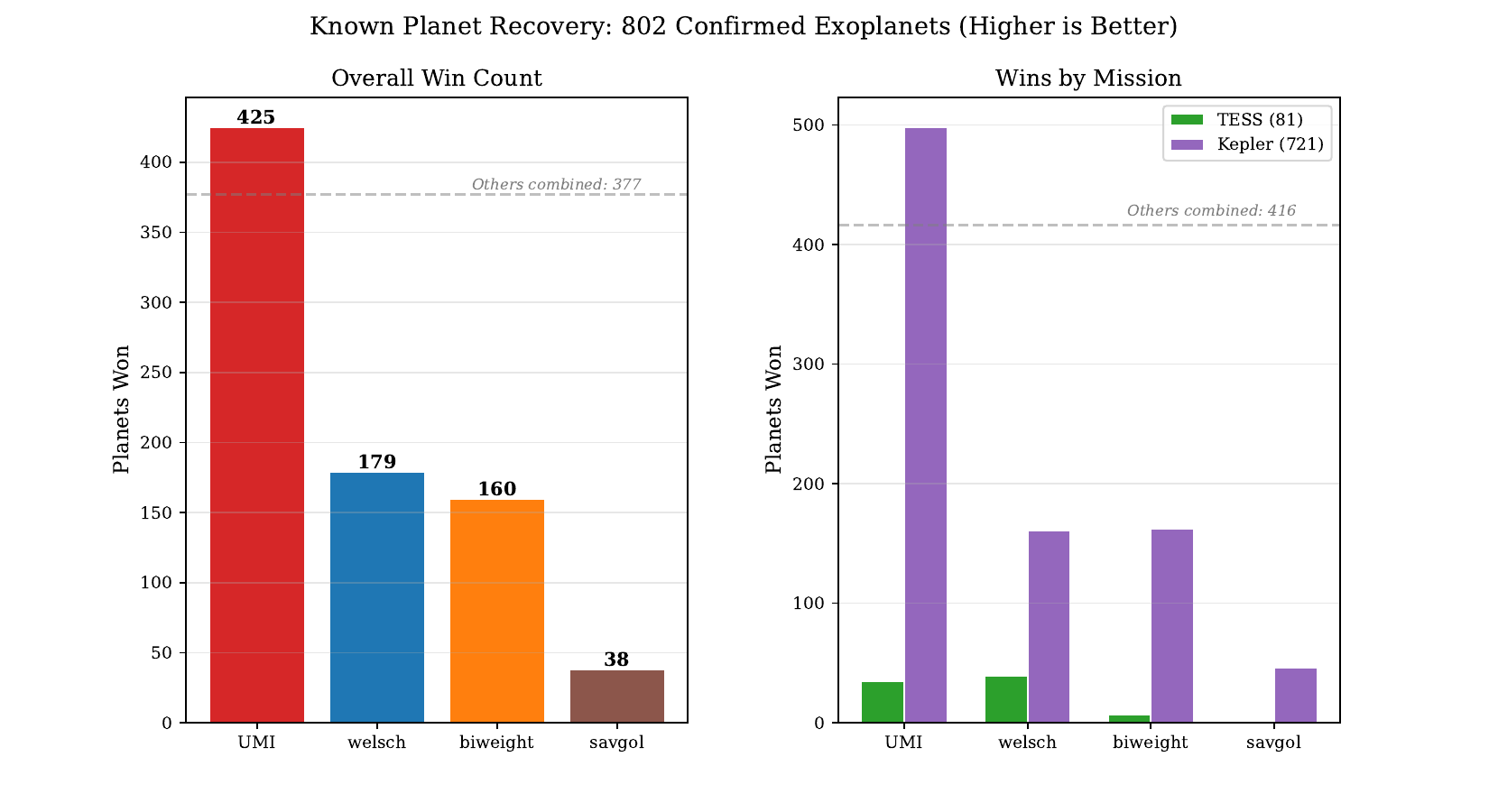}
\caption{Known planet recovery on 802 confirmed exoplanets (81 TESS,
721 Kepler). Left: overall win count, showing that UMI recovers more
planets (425) than all other methods combined (377). Right: wins broken
down by mission, showing that UMI dominates on Kepler while remaining
competitive on TESS. Higher is better.}
\label{fig:known_planets}
\end{figure*}

UMI achieves the best depth recovery for 425 of the 802 planets
(53\%), with the advantage concentrated on Kepler's higher-precision
data where UMI's win rate is 54\%. On TESS, UMI and Welsch produce
comparable results (35 vs 38 wins on 81 planets), consistent with
the regime analysis above: 62\% of confirmed TESS planets have
depths exceeding 10,000~ppm (1\%), where symmetric methods already
reject transit points effectively and UMI's asymmetric advantage
is minimal. These results are shown in Figure~\ref{fig:known_planets}.

\subsection{Multi-Mission Consistency}

\begin{figure*}[t!]
\centering
\includegraphics[width=\textwidth]{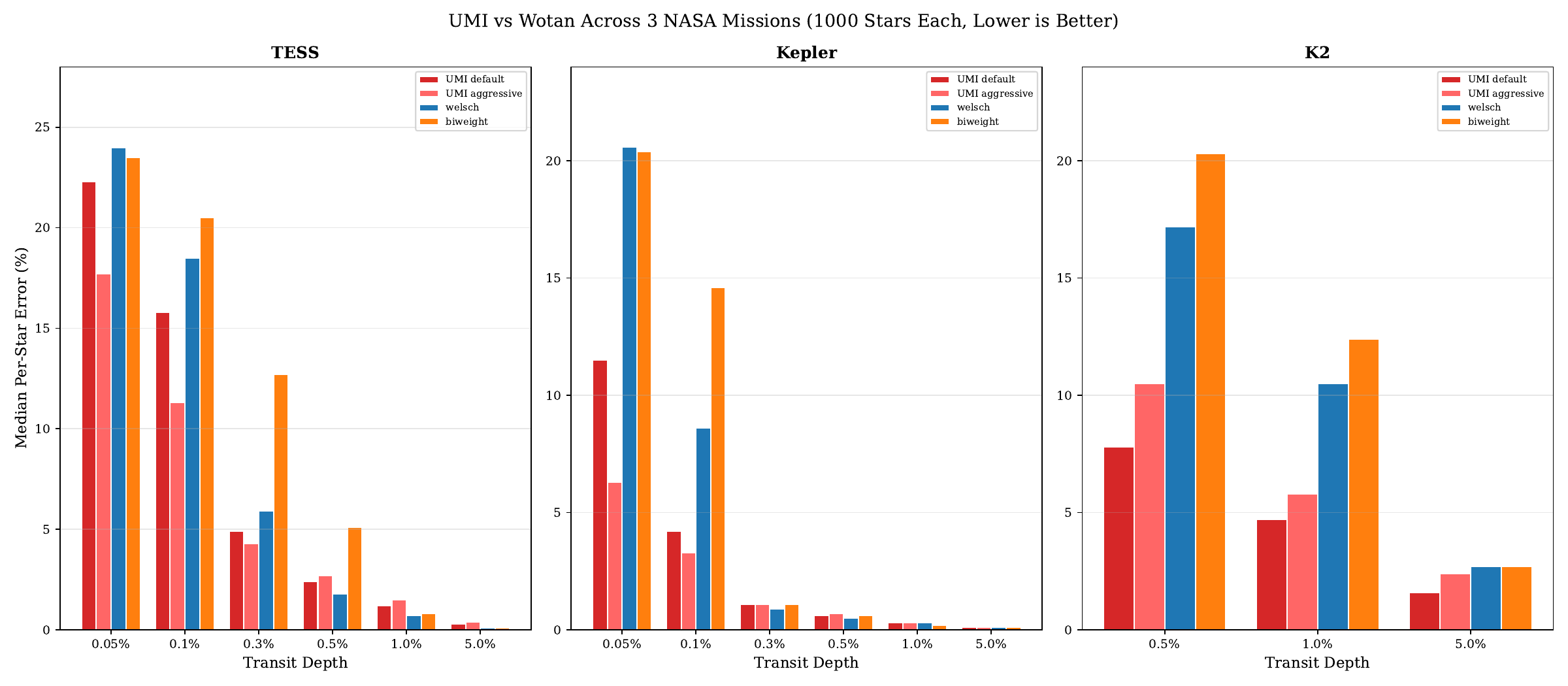}
\caption{UMI versus wotan across 3 NASA missions (1000 stars each).
UMI default and aggressive modes are compared against Welsch and
biweight at 6 depths for TESS and Kepler, and 3 depths for K2
(K2's higher noise floor renders the shallowest depths
uninformative; the 0.1\% K2 results exceed 400\% error for all
methods and are omitted for clarity).
Lower is better.}
\label{fig:multi_mission}
\end{figure*}

UMI was validated across three NASA missions: TESS \citep{ricker2015},
Kepler \citep{borucki2010}, and K2 \citep{howell2014}
(Figure~\ref{fig:multi_mission}). On 1000 Kepler Q5 stars at
0.1\% depth, UMI achieves 4.2\% error compared to 14.6\% for
biweight, a 71\% improvement. On K2 at 0.5\% depth, UMI achieves
7.8\% versus 20.3\% for biweight, a 62\% improvement.

\begin{figure}[t!]
\centering
\includegraphics[width=\columnwidth]{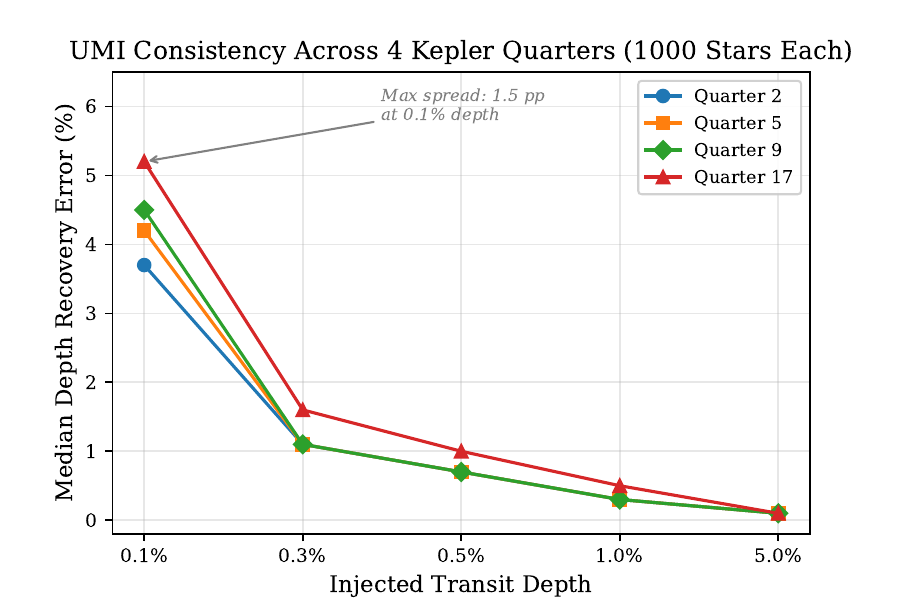}
\caption{UMI consistency across 4 Kepler quarters (Q2, Q5, Q9, Q17;
1000 stars each). The maximum spread is 1.5 percentage points at
0.1\% depth, demonstrating stable performance across different
observing epochs.}
\label{fig:kepler}
\end{figure}

Multi-quarter Kepler validation (Q2, Q5, Q9, Q17; 1000 stars each)
confirms consistency: 3.7 to 5.2\% error at 0.1\% depth across all
four quarters (Figure~\ref{fig:kepler}). This demonstrates that
UMI's performance is stable across different observing epochs and
is not an artifact of a particular quarter's systematics.

\subsection{Speed}

Benchmarked on an AMD Radeon RX 9060 XT (16~GB VRAM) with real
TESS sector~6 data (19,618 stars), UMI achieves 154 stars per
second for the full preprocessing pipeline, compared to 4.2 stars
per second for wotan with 12 CPU workers, a factor of 37 improvement
(Table~\ref{tab:speed}). A full TESS sector completes in 2.1 minutes.
Per-star detrending time is 3.4~ms compared to 234~ms for biweight,
a factor of 69 improvement (Figure~\ref{fig:speed}).

\begin{table}[h]
\centering
\caption{Processing speed comparison for full pipeline (quality
filtering, gap handling, detrending, normalization, windowing).}
\label{tab:speed}
\begin{tabular}{lrr}
\hline
Pipeline & Rate & Speedup \\
\hline
\texttt{wotan} biweight (12 CPU workers) & 4.2 stars/sec & 1$\times$ \\
\textbf{TorchFlat UMI (GPU)} & \textbf{154 stars/sec} & \textbf{37$\times$} \\
\hline
\end{tabular}
\end{table}

\section{Discussion}
\label{sec:discussion}

\subsection{Variable Stars}

The default $\alpha = 2.0$ introduces a systematic bias of $-451$~ppm
on quiet stars and approximately $-7240$~ppm on stars with intrinsic
photometric variability exceeding 1\%. This bias arises because the
asymmetric weighting interprets stellar variability dips as transit
signals and systematically excludes them from the trend estimate,
pulling the trend above the true mean. Users analyzing variable star
populations should set $\alpha = 1.0$, which restores symmetric
biweight behavior and reduces the bias to $-2$~ppm, at the cost of
losing the asymmetric depth-preservation advantage. A bias correction
lookup table indexed by variability amplitude is provided in the
software for intermediate cases.

\subsection{Long-Duration Transits}

With the default 0.5-day window ($W = 361$ for TESS), UMI
underestimates transit depth for transits longer than approximately
5 hours, because the window cannot fully bracket the transit and
treats parts of the ingress and egress as continuum. We recommend a
window length of at least 2 to 3 times the expected transit duration,
or \texttt{window\_length=1.5} days for transits up to 12 hours. This
limitation is fundamental to any sliding-window method and is not
specific to UMI.

\subsection{Recommended Parameter Settings}

Table~\ref{tab:params} summarizes the recommended parameter settings
for common use cases, based on the regime analysis in
Section~\ref{sec:validation}.

\begin{table}[h]
\centering
\caption{Recommended UMI parameters by use case.}
\label{tab:params}
\begin{tabular}{lcc}
\hline
Use case & $\alpha$ & $c$ \\
\hline
TESS planet search (default) & 2.0 & 5.0 \\
Kepler / high-precision data & 3.0 & 5.0 \\
Variable star surveys & 1.0 & 5.0 \\
Maximum shallow sensitivity & 10.0 & 2.5 \\
\hline
\end{tabular}
\end{table}

\subsection{Kepler Long-Cadence}

Kepler's 30-minute cadence produces a window size of only $W = 25$
points, compared to $W = 361$ for TESS 2-minute cadence. The minimum
segment length parameter is automatically scaled to $W/3$ to avoid
producing invalid output, but this means fewer points contribute to
each trend estimate. Despite this, UMI still outperforms biweight on
Kepler data (Section~\ref{sec:validation}).

\subsection{False Positive Rates}
\label{sec:false_positive}

While UMI preserves transit depth better than symmetric methods, the
$-451$~ppm bias could in principle affect downstream transit search
algorithms such as BLS \citep{kovacs2002} or TLS
\citep{hippke2019tls}. However, the bias is a constant offset applied
uniformly to all cadences, not a periodic signal. Transit search
algorithms detect periodic dips by phase-folding at trial periods, and
a constant offset shifts the entire baseline equally without creating
the periodic structure that triggers a detection. The bias therefore
affects absolute flux calibration but should not increase the false
positive rate. A systematic verification using matched-filter
detection on detrended light curves without injected transits would
confirm this reasoning and is planned for future work.

\subsection{Comparison with Gaussian Processes}

GP regression (e.g., \texttt{celerite}; \citealt{foremanmackey2017})
models the covariance structure of stellar variability and can produce
physically motivated noise models. However, GPs operate on a per-star
basis (minutes per star) and are unsuitable for survey-scale
preprocessing. UMI is designed for initial bulk detrending at
154~stars per second; GP modeling should be applied to individual
targets of interest for precise parameter estimation.

\subsection{Relation to \texttt{wotan}}

The UMI algorithm (asymmetric weight function and upper-RMS scale)
could be implemented as a new method within \texttt{wotan} itself,
running on CPU in the same sliding-window framework. This would
provide the accuracy improvement over the standard biweight at
comparable speed. However, the 69$\times$ detrending speedup
reported in this work comes specifically from the fused GPU kernel
in \texttt{torchflat}, which parallelizes hundreds of thousands of
window evaluations simultaneously and eliminates intermediate memory
allocations. A CPU implementation of UMI would be roughly 1.5 to
2$\times$ faster than the standard biweight (because upper-RMS
avoids the MAD sort), not 69$\times$.

We envision a natural division: \texttt{wotan} could offer UMI as a
CPU method for users who want the accuracy benefit without requiring
a GPU, while \texttt{torchflat} provides the GPU-accelerated
implementation for survey-scale processing.

\subsection{Scaling to Future Missions}

UMI's accuracy advantage over symmetric methods grows with
photometric precision. On TESS ($\sigma \approx 1000$~ppm), a 0.1\%
transit produces a residual of approximately $1\sigma$, which is
difficult for any method to distinguish from noise. On Kepler
($\sigma \approx 100$~ppm), the same transit is $10\sigma$, and the
asymmetric weight clearly identifies it as a downward outlier and
excludes it from the trend. This is reflected in our results: UMI's
improvement over biweight at 0.1\% depth grows from 23\% on TESS to
71\% on Kepler.

The upcoming PLATO mission \citep{rauer2014} is expected to achieve
noise levels of 30 to 50~ppm for bright stars, placing even shallow
transits at 20 to $30\sigma$. At these signal-to-noise ratios, the
asymmetric weight will assign near-zero weight to transit points,
providing nearly perfect depth preservation. Moreover, PLATO will
monitor hundreds of thousands of stars at 2-minute cadence, making
the 69$\times$ detrending speedup increasingly important for
survey-scale processing. UMI is designed to scale to this regime.

\section{Conclusion}
\label{sec:conclusion}

We have presented UMI, a modification of the Tukey bisquare
M-estimator that exploits the physical one-sidedness of transit
signals through an asymmetric weight function and a one-sided scale
estimator. These are simple changes to the standard biweight that
require no additional computational cost but yield measurable
improvements in transit depth preservation, particularly at depths
above the photometric noise floor where the asymmetric weighting can
distinguish transit dips from photometric noise.

The depth regime characterization presented in
Tables~\ref{tab:tess}--\ref{tab:kepler} shows that the advantage
is not uniform: at exomoon-scale depths ($\leq 50$~ppm), all
sliding-window methods are noise-limited and produce comparable
results. UMI's contribution is concentrated at planet-scale depths
(200--1000~ppm on TESS, 100--1000~ppm on Kepler), which is the
regime most relevant to current and upcoming transit surveys.

The fused GPU kernel implementation demonstrates that robust
statistical methods can be accelerated to survey scale without
sacrificing accuracy. As transit surveys grow in volume with
missions such as PLATO \citep{rauer2014}, efficient detrending will
become increasingly important.

Future work will focus on three areas: (1) a systematic false
positive analysis to quantify the effect of asymmetric bias on
downstream transit detection, (2) adaptive asymmetry selection
based on per-star noise properties, and (3) hierarchical multi-scale
extensions that incorporate global variability awareness while
maintaining GPU-speed processing.

TorchFlat is open-source (MIT license) and available via
\texttt{pip install torchflat}. Source code and all validation results
are at \url{https://github.com/omarkhan2217/TorchFlat}.

\begin{acknowledgments}
We thank the TESS and Kepler teams for making their data publicly
available through MAST. This work made use of \texttt{PyTorch}
\citep{pytorch2019}, \texttt{wotan} \citep{hippke2019}, and
\texttt{lightkurve} \citep{lightkurve2018}.
\end{acknowledgments}

\bibliography{references}

@article{hippke2019,
    author = {Hippke, Michael and David, Trevor J. and Mulders, Gijs D. and Heller, Ren{\'e}},
    title = {Wotan: Comprehensive Time-series De-trending in Python},
    journal = {The Astronomical Journal},
    volume = {158},
    number = {4},
    pages = {143},
    year = {2019},
    doi = {10.3847/1538-3881/ab3984}
}

@article{foremanmackey2017,
    author = {Foreman-Mackey, Daniel and Agol, Eric and Ambikasaran, Sivaram and Angus, Ruth},
    title = {Fast and Scalable Gaussian Process Modeling with Applications to Astronomical Time Series},
    journal = {The Astronomical Journal},
    volume = {154},
    number = {6},
    pages = {220},
    year = {2017},
    doi = {10.3847/1538-3881/aa9332}
}

@inproceedings{pytorch2019,
    author = {Paszke, Adam and Gross, Sam and Massa, Francisco and others},
    title = {PyTorch: An Imperative Style, High-Performance Deep Learning Library},
    booktitle = {Advances in Neural Information Processing Systems 32},
    year = {2019},
    pages = {8024--8035}
}

@misc{lightkurve2018,
    author = {{Lightkurve Collaboration}},
    title = {Lightkurve: Kepler and TESS time series analysis in Python},
    year = {2018},
    howpublished = {Astrophysics Source Code Library, ascl:1812.013}
}

@article{hoare1961,
    author = {Hoare, C. A. R.},
    title = {Algorithm 65: Find},
    journal = {Communications of the ACM},
    volume = {4},
    number = {7},
    pages = {321--322},
    year = {1961},
    doi = {10.1145/366622.366647}
}

@book{tukey1977,
    author = {Tukey, John W.},
    title = {Exploratory Data Analysis},
    publisher = {Addison-Wesley},
    year = {1977}
}

@book{hampel1986,
    author = {Hampel, Frank R. and Ronchetti, Elvezio M. and Rousseeuw, Peter J. and Stahel, Werner A.},
    title = {Robust Statistics: The Approach Based on Influence Functions},
    publisher = {John Wiley \& Sons},
    year = {1986}
}

@article{rousseeuw1993,
    author = {Rousseeuw, Peter J. and Croux, Christophe},
    title = {Alternatives to the Median Absolute Deviation},
    journal = {Journal of the American Statistical Association},
    volume = {88},
    number = {424},
    pages = {1273--1283},
    year = {1993},
    doi = {10.1080/01621459.1993.10476408}
}

@article{ricker2015,
    author = {Ricker, George R. and Winn, Joshua N. and Vanderspek, Roland and others},
    title = {Transiting Exoplanet Survey Satellite ({TESS})},
    journal = {Journal of Astronomical Telescopes, Instruments, and Systems},
    volume = {1},
    pages = {014003},
    year = {2015},
    doi = {10.1117/1.JATIS.1.1.014003}
}

@article{borucki2010,
    author = {Borucki, William J. and Koch, David and Basri, Gibor and others},
    title = {Kepler Planet-Detection Mission: Introduction and First Results},
    journal = {Science},
    volume = {327},
    number = {5968},
    pages = {977--980},
    year = {2010},
    doi = {10.1126/science.1185402}
}

@article{howell2014,
    author = {Howell, Steve B. and Sobeck, Charlie and Haas, Michael and others},
    title = {The {K2} Mission: Characterization and Early Results},
    journal = {Publications of the Astronomical Society of the Pacific},
    volume = {126},
    number = {938},
    pages = {398},
    year = {2014},
    doi = {10.1086/676406}
}

@article{kovacs2002,
    author = {Kov{\'a}cs, G{\'e}za and Zucker, Shay and Mazeh, Tsevi},
    title = {A box-fitting algorithm in the search for periodic transits},
    journal = {Astronomy \& Astrophysics},
    volume = {391},
    pages = {369--377},
    year = {2002},
    doi = {10.1051/0004-6361:20020802}
}

@article{hippke2019tls,
    author = {Hippke, Michael and Heller, Ren{\'e}},
    title = {Optimized transit detection algorithm to search for periodic transits of small planets},
    journal = {Astronomy \& Astrophysics},
    volume = {623},
    pages = {A39},
    year = {2019},
    doi = {10.1051/0004-6361/201834672}
}

@article{akeson2013,
    author = {Akeson, R. L. and Chen, X. and Ciardi, D. and others},
    title = {The {NASA} Exoplanet Archive: Data and Tools for Exoplanet Research},
    journal = {Publications of the Astronomical Society of the Pacific},
    volume = {125},
    number = {930},
    pages = {989},
    year = {2013},
    doi = {10.1086/672273}
}

@article{cleveland1979,
    author = {Cleveland, William S.},
    title = {Robust Locally Weighted Regression and Smoothing Scatterplots},
    journal = {Journal of the American Statistical Association},
    volume = {74},
    number = {368},
    pages = {829--836},
    year = {1979},
    doi = {10.1080/01621459.1979.10481038}
}

@article{rauer2014,
    author = {Rauer, H. and Catala, C. and Aerts, C. and others},
    title = {The {PLATO} 2.0 mission},
    journal = {Experimental Astronomy},
    volume = {38},
    pages = {249--330},
    year = {2014},
    doi = {10.1007/s10686-014-9383-4}
}
\bibliographystyle{aasjournal}

\end{document}